# Decoherence and wave function collapse

Roland Omnès*
Laboratoire de Physique Théorique[#]
Université de Paris XI, Bâtiment 210, F-91405 Orsay Cedex, France

## Abstract

The possibility of consistency between the basic quantum principles of quantum mechanics and wave function collapse is reexamined. A specific interpretation of environment is proposed for this aim and applied to decoherence. When the organization of a measuring apparatus is taken into account, this approach leads also to an interpretation of wave function collapse, which would result in principle from the same interactions with environment as decoherence. This proposal is shown consistent with the non-separable character of quantum mechanics.

[

## 1. Introduction

The problem of wave function collapse stands still at the center of interpretation in quantum mechanics [1]. The only universal –or at least almost universal– agreement on it is that this collapse, or reduction, or the uniqueness of macroscopic reality, cannot be a consequence of quantum dynamics: Schrödinger ruled this idea out already because it would imply, at face value, two incompatible dynamics [2]. After mentioning the attractiveness of their conciliation, Wigner gave a careful proof of its impossibility in what may be called a no-go theorem [3], which was more recently confirmed on different assumptions by Bassi and Ghirardi [4].

Many answers to the problem of collapse, often deep and imaginative, have been envisioned. They proposed various completions, revisions or extensions of quantum mechanics, with much divide opinions on their respective values [5] . They split the scientific community into different schools and, according to their various viewpoints; every limited advance in interpretation soon becomes a subject of controversy.

There were nevertheless several such advances, although they shed no light on collapse. One may mention for instance decoherence [6-10], which explained the observational absence macroscopic interferences, such as the addition of the states "dead" and "alive" in the case of Schrödinger's cat. Decoherence was also confirmed by experiment [11]. The emergence of classical dynamics at a macroscopic scale was derived mathematically from quantum dynamics [12], and also less formally from decoherence [13] Consistent histories clarified the language and logic of quantum mechanics [14-16], and removed many paradoxes [17].

A common feature in these developments was their direct derivation from the basic quantum principles, suggesting the possibility of still undiscovered non-trivial other consequences in these principles. Such an approach, admittedly philosophic, would imply that the principles contain in germ all the concepts allowing their own interpretation and it suggests to reconsider in this light the problem of collapse and more widely a possible consistency or of quantum dynamics and its completeness. This is the topic of the present work. It will mainly proceed trough a critique of measurement theory rather than a search for a mechanism of collapse .It will not be on the other hand a complete critique of measurement theory, which exists in good books [1, 5, 18, 19], and will be devoted on the contrary to a unique question: "Could collapse emerge from quantum dynamics?" One will try to find whether the answer is negative, as usually believed, or if a reexamination can open a suggestive explanation of collapse and the uniqueness of reality.

One will take for granted the principles of quantum mechanics, which can be summarized for the present purpose by an axiom assuming that an isolated physical system is associated with a definite Hilbert space $E$ and a set of observable involving especially a Hamiltonian $H$, the wave functions, or vectors in $E$, evolving under the unitary operator $U(t) = \exp(-iHt/\hbar)$.

One may notice however that the notion of "physical system", which is clear in classical physics, is somewhat fuzzy in quantum mechanics [19]. One knows since a long time that no local system can be exactly isolated [20], so that the axiom is basically a hypothesis in Poincaré' sense [21]. As such, its use obeys methodological rules. When dealing with an elementary object (an atom for instance) and a definite kind of properties (*e.g.*, the energy spectrum), one can check that a lack of isolation does not spoil the conclusions by making simple estimates. In a more complex situation (dealing for instance with a solid or a chemical reaction), one can proceed in the same way, assume first that the solid is isolated and then check that the conclusions are sensible, but always for a specific type of object and a specific type of property in which one is interested..

In the case of a measurement, the first step of this approach consists in considering the measuring apparatus $A$ and the measured object $m$ as two interacting subsystems of an isolated system $A + m$ [2]; The most essential consequences of the analysis can be obtained in a case where the measured observable $Z$ is two-valued with eigenvectors $|1\rangle$ and $|2\rangle$. One defines then formally $A$ as a measuring device for the observable $Z$ by imposing reliability conditions as follow: If the initial state of $m$ is $|1\rangle$ (or respectively $|2\rangle$), a pointer in $A$ shows after measurement a definite position $X = x_1$ (respectively $x_2$) [2-4, 22]. For convenience, one can also restrict attention to the case where the states $|1\rangle$ or $|2\rangle$ of $m$ are conserved after these measurements.

If the initial state of $A$ can be described by a wave functions or a state vector $|A_0\rangle$ and the final state is $|A_1\rangle$ (respectively $|A_2\rangle$) when $|1\rangle$ (respectively $|2\rangle$) is measured ($|A_2\rangle$, the basic axiom implies that at a time $t_f$ marking the end of the measurement, one has

$$U_{A+m}(t_f)|A_0\rangle|1\rangle = |A_1\rangle|1\rangle, \quad U_{A+m}(t_f)|A_0\rangle|2\rangle = |A_2\rangle|2\rangle. \tag{1.1}$$

If the initial state of *m* is a superposition

$$|s\rangle = c_1|1\rangle + c_2|2\rangle, \tag{1.2}$$

unitarity implies that the final state of *A* + *m* is

$$c_1|A_1\rangle|1\rangle + c_2|A_2\rangle|2\rangle.. \tag{1.3}$$

This prediction is contradicted however by experiment, except in the case of models of measurement (particularly in quantum optics) in which *A* and *m* are simple systems, initially in pure states, and measurement is a quantum jump [23].

Decoherence entered essentially measurement theory as the next step in methodology verifying that the conclusions obtained for an isolated system remain valid for a non-isolated (open) system. An environment *e* of the apparatus *A* was introduced and the *A-e* interactions were investigated. A remarkable consequence was an explanation for the absence of macroscopic interferences from a superposition, such as the state "dead + alive" in the case of Schrödinger's cat. The conclusions did not affect however the main contradiction with observation, namely the final values of the squared amplitudes of the two terms in Eq.(1.3):

$$p_1 = |c_1|^2, \quad p_2 = |c_2|^2, \tag{1.4}$$

which mean that the two positions of the pointer are still present at the end of the measurement, although the two events are now mutually exclusive ("either or") and do not interfere.

A critique of the assumptions of these standard approaches and of their interpretations will be proposed in this paper. Section 2 deals with the meaning and use of wave functions, and with the status of environment in respect to the basic axiom. A new interpretation of decoherence is proposed in Section 3 and its main result is a critique of the assumptions in no-go theorems, particularly the reliability conditions. Section 4 points out two significant features of a real measuring apparatus, namely its organization and the transport of incoherence in it. The prediction (1.4) is shown questionable and, on the contrary, fluctuations in $p_1$ and $p_2$ are found possible, or at least worth more investigation. Section 5 recalls how such fluctuations could lead to collapse. These considerations are kept however at the level of principles and no attempt is made in this work to obtain an explicit local mechanism of collapse with quantitative predictions, because of explicit technical difficulties. The interest of pursuing nevertheless this research is stressed in Section 6, where one shows that such a local mechanism does not disagree with the non-separability of quantum mechanics.

**2. Wave functions and environment**

*Wave functions*

As indicated in Section 1, a methodological relation between quantum measurements and the basic principles must start from the case of an isolated system *A* +*m*. The relevance of wave functions is less obvious, because the state of this macroscopic system is certainly not defined by a unique wave function. The right approach is rather to define the notion of state from the principles, as done by Von Neumann [22,12]: After associating the physical properties of the system with projection operators in its Hilbert space, the state is defined as a datum attributing a probability to every property. Considering logically compatible properties as associated with commuting projection operators, some basic axioms of logic imply that the state is necessarily associated with a density matrix $\rho_{A+m}$ [24].

One notices for future use that when a system *S* consists of two subsystems *A* and *B*, the same logical considerations imply that the density matrix of the subsystem *A* must defined by a trace over the degrees of freedom of the subsystem *B*, namely

$$\rho_A = Tr_B(\rho_S). \tag{2.1}$$

Coming back then to the standard considerations in Section 1, one notices that they rely heavily on wave functions so that, to give them a precise meaning, one must define these wave functions. The only procedure starting from the basic axiom consists then in considering these wave functions as eigenfunctions (or eigenvectors) of $\rho_{A+m}$. But methodology asks that one checks later on that the conclusions are insensitive to a lack of exact isolation, and one may guess that difficulties will appear: One knows that the eigenvectors s of an operator are extremely sensitive to perturbations when the eigenvalues of this operator are very close, and that will turn out to the situation when one will examine it in Section 3. In other words, the conclusions about measurements resulting from a use of wave functions are presumably very sensitive to external perturbations and they must be considered with much care.

The conditions of reliability defining a measuring apparatus raise also some questions. The independence of the apparatus *A* and the object *m* before measurement means that *A* and *m* are then independent systems, associated with density matrices $\rho_A$ and $\rho_m$. Independence implies then that $\rho_{A+m} = \rho_A \otimes \rho_m$. Once again, a density matrix $\rho_m$ comes prior to its eigenvectors: The two vectors $|1\rangle$ and $|2\rangle$ do not enter in the conditions of reliability as mathematical vectors but as two states that are measured. As such, one is not measuring a state vector $|1\rangle$ or $|2\rangle$, but a state happening to be pure and described either by $\rho_m = |1\rangle\langle 1|$ or $|2\rangle\langle 2|$. The physical meaning of the reliability conditions rests therefore on a prior condition for the initial state, which can be justified only by Dirac's methodology: A state is known when it comes from some measurement or preparation [25]. This point may look accessory at first sight, but it will have significant consequences.

*Decoherence and environment*

From the standpoint of methodology, decoherence theory can be considered as the second step when corrections are introduced to the assumption of isolation for the system *A* + *m* and its environment must enter. Using only one example for the sake of brevity, one will consider an environment *e* consisting of an atmosphere of molecules *M*, these molecules being in practice structureless and spinless. Decoherence is then a consequence of the collisions of these molecules on *A*. According to an usual interpretation [13, Chapter 1], these collisions produce an entanglement between the wave functions of *A* and those of *e*, together with random phase shifts resulting from scattering [10]. The growth of this entanglement with wider region in the environment generates a decrease in the non-diagonal elements of the reduced density matrix $\rho_r$; which represents the pointer position. One gets typically

$$\rho_r(x_1,x_2;t) \approx \rho_r(x_1,x_2;0)\exp(-t/\tau_d), \tag{2.1}$$

where $\tau_d$ is a (generally very short) decoherence time. This exponential damping implies a removal of macroscopic interferences, but the diagonal elements of $\rho_r$ are unaffected and the unchanged probabilities $p_1$ and $p_2$ of the two channels are still given by Eq.(1.4).

Contrary to the concept of system, which is demanding from a physical standpoint and sharp from a mathematical one, the notion of environment is fuzzy in both aspects. That does not mean that it is worthless, but a serious difficulty with this notion is to assign it a boundary to $e$ so as to make it a system: A circumscribed environment has always a wider environment. This difficulty is sometimes circumvented by taking the whole universe outside $A + m$ as the environment and assuming a pure state of the universe [13, 26], but this way out has drawbacks, not even mentioning that it extends the axioms of quantum mechanics far wider than their established domain and precisely to bypass their limitations.

The main drawback arises from the conditions of reliability when they are associated with explicitly prepared states $|1\rangle\langle1|$ or $|2\rangle\langle2|$. The existence of a preparing apparatus, which does not disappear completely and keeps some memory of the preparation, belongs then also to the wave function of the universe and Eqs.(1.1-3) become absurd: The sum (1.3), now understood as valid with states $|A\rangle$ in the universe, involves in principle an apparatus preparing $|s\rangle\langle s|$, whereas (1.3) is a sum of wave functions of the universe in which the states $|1\rangle\langle1|$ and $|2\rangle\langle2|$ are prepared quite differently. This sounds at least puzzling.

The question of circumscribing $e$ by a boundary is also not harmless. Once again, when examining no-go theorems, one must look at wave functions and in the present case at wave functions of $A+m+e$. But however far the boundary of $e$ is pushed, the eigenvector of $\rho_{A+m+e}$ remain extremely sensitive to unavoidable interactions with a still wider environment.

This point is ignored in practical applications of decoherence theory, where $e$ is treated as a system consisting of a definite collection of particles or of harmonic oscillators. There is much sense in this standpoint when one considers it as a check of decoherence theory, because the damping (2.1) of non-diagonal elements of $\rho_r$ and the explicit expression of $\tau_d$ depend only on the properties of $e$ in the immediate vicinity of $A$. Only the mass of molecules $M$, their space density and their temperature are needed. The conclusions about decoherence are therefore certainly robust and insensitive to refinements trying to circumscribe the fuzziness of $e$.

But the interpretation of decoherence as a growth in entanglement has some drawbacks. It relies on the idea that after colliding with $A$, the outgoing molecules suffer collisions with other molecules in $e$ and extend their entanglement to larger and larger distances. There is no end to this growth and some authors assume even that this entanglement can extend to the whole universe, leading finally to some version of Everett's interpretation [13, 26]. Whatever the suggestive value of this interpretation, one may feel uneasy about it: Why should the whole universe be an actor in an event as inconspicuous at a cosmological scale as a unique local measurement?

One may have restrictions on this interpretation and look for a more convenient one. The interpretation that will be used here consists simply in recognizing that an environment can never be conceived exactly as a quantum system with the formal mathematical machinery of Hilbert space and density matrix. This machinery is valuable for decoherence in a restricted sense, *i.e.*, as far as the damping of non-diagonal elements in the reduced density matrix for a pointer is concerned, as one already explained. It is questionable however for other predictions such as the conservation of the channel probabilities (1.4), if only because, experimentally, this conservation is satisfied when $A$ and $m$ are in a pure state [23], but never otherwise. In other words, decoherence in a strict sense is observed experimentally [11], but the conservation of probabilities is not. This is enough to shed doubt on the validity of the approach itself, or at least on some of its aspects.

Rather than using fuzzy interpretations of the environment, it seems better to look only at what is actually responsible for decoherence, namely the existence of nearby molecules and their statistical distribution. Their collisions with *A* are quantum events and the results of decoherence calculations depend only on the quantum description of each colliding molecule (or of the behavior each harmonic oscillator), and non-diagonal interference effects disappear only because of an addition of these individual effects. One needs nowhere an explicit and detailed density matrix $\rho_e$ or its entangled wave functions. This is why one will use here a model of environment involving only the arrival of molecules *M* on *A* as a statistical effect, described by probability calculus and not necessarily by a full quantum theory requiring an explicit density matrix for *e*. The quantum description will be restricted to an account of each collision, with no consideration for what happens to this molecule after it collided. For convenience, one will call it a "probabilistic environment" and one will consider in the next sections how this empirical interpretation compares with previous ones.

**3. Decoherence and no-go theorems**

The first aim of this section is to compare the theory of decoherence in the framework of a probabilistic environment with the usual theory where *e* is considered as a quantum system. Since the whole theory must be derived in a new framework, its development is necessarily rather long and technical. They will be performed in the case where *A* is considered globally and characterized only by the position *X* of a pointer, as usual. The results of this approach will be found identical with previous ones, except for an essential property of the wave functions of *A*, defined as eigenfunctions of $\rho_A$: they do not evolve unitarily and this property points out a loophole in the assumptions of no-go theorems.

*External collisions*

When the apparatus *A* is isolated, one can define its wave functions as the eigenfunctions of $\rho_A$ at a definite time together with the corresponding eigenvalues through

$$\rho_A = \sum_k p_k |A,k\rangle\langle A,k|, \qquad (3.1)$$

The index *k* is discrete because *A* is bounded and one will say that it defines a sector of $\rho_A$. The technical apparel of decoherence theory will be first introduced here in the case of an apparatus *A*, before measurement, when there is yet no entanglement with *m*, because some effects of external collisions will be found already significant in this case and will become essential in the next section.

One will assume that *A* is in thermal equilibrium so that a state $|A,k\rangle$ is an eigenvector of the Hamiltonian $H_A$ with an eigenvalue $E_k$ (usually degenerate). One is interested in the change of $\rho_A$ when an external molecule *M* collides with *A*. From the standpoint of the basic quantum axiom and of a probabilistic interpretation of environment, this means that one is considering an event where *M* is initially in an eigenstate $|q\rangle$ of momentum and collides with *A* in the state $\rho_A$. In other words, one is considering the quantum system *A+M* in the initial state $|q\rangle\langle q| \otimes \rho_A$.

Using a shortened notation where a state vector $|A,k\rangle|q\rangle$ is written $|kq\rangle$, one can use collision theory where the *S*-matrix is written as $I - iT$ and *T* is the so-called collision matrix [27]. One has then

$$S|kq\rangle = (1 - \sum_{k'q'}|\langle k'q'|T|kq\rangle|^2)^{1/2}|kq\rangle - i\sum_{k'q'}\langle k'q'|T|kq\rangle|k'q'\rangle. \qquad (3.2)$$

The notation has been again shortened and is understood in the general case of an initial state *i* and a final state *f* with a notation where

$$S_{fi} = \delta_{fi} - 2\pi i \delta(E_f - E_i)\langle f|T|i\rangle \text{ and } \sum_{q'} = \int (2\pi)^{-3} d^3 q', . \quad (3.3)$$

If the probability of occurrence of the incoming state $|q\rangle\langle q|$ is $\varepsilon$, the change $\delta\rho_A$ in $\rho_A$ after the collision is given, according to Eq.(2.1) by

$$\delta\rho_A = -\varepsilon \sum_{kk'q'} p_k |k\rangle\langle k| |\langle k'q'|T|kq\rangle|^2$$
$$+ \varepsilon \sum_{kk'k''q'} |k'\rangle\langle k''| \cdot \{\langle k'q'|T|kq\rangle\langle k''q''|T|kq\rangle^*\} p_k. \quad (3.4)$$

As could be expected, one notices that:

$$Tr(\delta\rho_A) = 0; \qquad \delta\rho_A = \delta\rho_A^\dagger. \quad (3.5)$$

The use of a plane wave $|q\rangle$ is inconvenient however because it implies a complete indeterminacy in the time of collision, which is however essential in decoherence. One must therefore *M* in space at some earlier time and this is best obtained by using a normalized wave coco packet $G(q, x)$ ([28], Chapter 8, [29]). One will suppose this packet Gaussian, spherical, with average momentum *q* and a definite average position *x* in the vicinity of *A* at some definite time. The time of arrival of *M* on *A* is then rather well defined.

The probability for a transition $k \rightarrow k'$ under the collision is then

$$P(kq_{in} \rightarrow k') = \int d^3q |G(q)|^2 (2\pi)^{-3} d^3q' (2\pi) \delta(E_{k'} + E_{q'} - E_k - E_q)|\langle k'q'|T|kq\rangle|^2, \quad (3.6)$$

the notation $q_{in}$ reminding that one is now dealing with a wave packet. This probability is related to the cross section by

$$P(kq_{in} \rightarrow k') = \sigma(kq_{in} \rightarrow k')/2\pi L^2, . \quad (3.7)$$

where *L* is the uncertainty in position of the wave packet and the cross section is given by

$$\sigma(kq_{in} \rightarrow k') = \int d^3q |G(q)|^2 (2\pi)^{-3} d^3q' (2\pi) \delta(E_{k'} + E_{q'} - E_k - E_q)$$
$$\times |\langle k'q'|T|kq\rangle|^2 / v. \quad (3.8)$$

where *v* denotes the velocity of *M*. Simple geometric considerations show that the ratio $\sigma(kq_{in} \rightarrow k')/L^2$ does not depend on the size *L* of the wave packet and one has furthermore

$$\sum_{k'} \sigma(kq_{in} \rightarrow k') = 2\pi L^2 \quad (3.9)$$

These properties are easily understood when one thinks of a localized particle hitting a large wall (which represents a part of the boundary of *A*, seen from a small scale). In practice, the most convenient expression for this size *L* is $L = n^{-1/3}$, where *n* is the space density of molecules in the environment.

The first term in Eq.(3.4) becomes then

$$-\varepsilon \sum_{kk'q'} p_k |k\rangle\langle k| |\langle k'q'|T|kq\rangle|^2 \rightarrow -(\varepsilon/2\pi L^2) \sum_{kk'} |k\rangle\langle k| p_k \sigma(kq_{in} \rightarrow k'). \quad (3.10)$$

The second term, which involves products of scattering amplitudes, is less transparent and useful only for qualitative considerations so that it is not written down explicitly.

One did not yet take account of the incoherence between different sectors, which is essential for the status of no-go theorems. In every sector, the representative vector $|Ak,t\rangle$ is only defined up to an arbitrary phase (one must consider it as a ray) One therefore introduces explicit random phases in the state vectors through

$$|k\rangle \to \exp(i\alpha_k)|k\rangle . \quad (3.11)$$

Eq.(3.1) becomes then

$$Se^{i\alpha_k}|kq\rangle = (1 - \sum_{k'q'}|\langle k'q'|T|kq\rangle|^2)^{1/2} e^{i\alpha_k}|kq\rangle$$
$$-i\sum_{k'q'}\langle k'q'|T|kq\rangle e^{i(\alpha_k - \alpha_{k'})} e^{i\alpha_{k'}}|k'q'\rangle. \quad (3.12)$$

One can interpret physically the last term as meaning that under a transition $k \to k'$, the sector $k$ injects into the sector $k'$ an incoherent addition with a random phase $\alpha_k - \alpha_{k'}$. Eq.(3.5) becomes then

$$(\delta\rho_A)_{k'k} = -\varepsilon\delta_{k'k}\sum_{k''q'} p_k |\langle k''q'|T|kq\rangle|^2$$
$$+\varepsilon\sum_{k''q'}\langle k'q'|T|k''q\rangle\langle kq'|T|k''q\rangle^* e^{i(\alpha_{k'}-\alpha_k)} p_{k''}. \quad (3.13)$$

*A loophole in no-go theorems*

The present interpretation of a probabilistic environment implies a significant consequence: In its framework, the evolution of an open system is not unitary. Before considering the signification and extent of this assertion, one will consider how it comes out, already, when no measurement is yet taking place and $A$ is an open system independent of $m$.

In that case, the density matrix $\rho_A$ of $A$ after a collision is defined according to Eq.(2.1) as the trace of $\rho_{A+M}$ over the outgoing states of $M$. The definition of wave functions of $A$ as eigenfunctions of $\rho_A(t)$ implies then that the new wave function in Sector $k$ becomes the corresponding perturbed eigenfunction of $\rho_A + \delta\rho_A$, with $\delta\rho_A$ given by Eq.(3.13). If perturbation calculus were valid, one would get at leading order in $\varepsilon$:

$$\delta p_k = \varepsilon\sum_{k'\neq k,q'}\{-p_k|\langle k'q'|T|kq\rangle|^2 + p_{k'}|\langle kq'|T|k'q\rangle|^2\}. \quad (3.14a)$$

$$\delta|k\rangle_\varphi = \varepsilon\sum_{k'k''q'}\langle kq'|T|k''q\rangle\langle k'q'|T|k''q\rangle^* |k'\rangle_\varphi \times [p_{k''}e^{i(\alpha_k-\alpha_{k'})}/(p_k - p_{k'})]. \quad (3.14.b)$$

The validity of perturbation calculus is questionable however, in view of the smallness of some denominators $p_k - p_{k'}$, but some qualitative features can be drawn anyway The main one is that the wave functions do not evolve unitarily. Every sector suffers continuously from a loss in norm and, simultaneously, receives incoherent injections from other sectors. This violation of unitarity, which is not negligible in principle, could be described as a source of nonlinearity in the Schrödinger equation, but its origin lies elsewhere: It boils down to the extraction of $\rho_A$ from $\rho_{A+M}$, which ignores the Schrödinger-Von Neumann equation

$$i(d/dt)\rho_{A+M} = -[H_{A+M}, \rho_{A+M}] \quad (3.15)$$

In that sense, this violation is general when wave functions are defined as eigenfunctions of the density matrix for a subsystem and it is not limited to the case of a probabilistic environment. Bassi and Ghirardi were aware of this difficulty, although they supposed the wave functions well defined by themselves and did not consider the effect of environment before measurement [4]. The difficulty arose then from the fuzziness of environment and they found only an issue through a selection of some properties of the environment having in principle relevance for the measurement ( they ignored preparation among these relevant properties).

In the present case, when measurement has begun, the same violation of unitarity occurs for $\rho_{A+m}$ when the state $|1\rangle\langle 1|$ or $|2\rangle\langle 2|$ is measured, so that when stating the conditions of reliability, Eqs.(1.1) are not valid. The same is true for the expression (1.3) of the final state when a pure state in superposition is measured. The main foundation of no-go theorems, which is the unitary evolution of wave functions, becomes therefore invalid.

What does it mean? It would be out of place to start here a controversy about the suitable framework for an account of environment. One would have expected however a strong robustness of the arguments establishing that quantum dynamics and collapse are incompatible. They should have been insensitive to a choice of interpretation, under the condition that this choice respects the axioms of quantum dynamics, but there is no such insensitivity. Moreover, a weak link appears in the use of a basic concept since it is concerned with unitarity. This means a loophole in the assumptions of no-go theorems, or at least a lack of robustness needing amendment. It does not mean on the other hand that the main conclusion of these theorems is automatically invalid. One cannot yet say that collapse is compatible with quantum mechanics, but it could be and this question will be considered in the next section.

*Decoherence*

Let one come back to decoherence. According to a probabilistic description of environment, the collisions of molecules on *A* are treated like quantum events with probabilistic initial conditions. Quantum wave packets arrive at various times from various directions and hit *A* (for instance on a solid boundary) at various places. One can then estimate some orders of magnitude if one takes into account the exponential form of thermal energy distribution for the molecules and the supposed Gaussian shape of wave packets. When taking an atmosphere under standard conditions as nearby environment and considering something between exp(-100) and exp(-10) as a good indicator of smallness, one finds that the size *L* of wave packets cannot be much smaller than $10^{-7}$ cm. This means that the interval of time during which this wave packet interacts with the boundary of *A* is of the order of $10^{-12}$ s, which is rather long when compared to the time scale $\tau_d$ of decoherence (of order $10^{-24}$ s if the size of *A* is in the centimeter range.

In these conditions, one may consider the action of *e* as a random arrival of molecules on *A*. Every molecule is described in a quantum way by a density matrix $|G\rangle\langle G|$ involving some wave packet *G*. The distribution of the center of this packet and of its average momentum can be considered on the other hand as classical. One is interested in the evolution of the density matrix $\rho_A$ and one canobtain it from the change occurring in $\rho_A$ during a small time interval $\delta t$. To do so, one considers its value $\rho_A(t)$ at some time *t*. For a given probabilistic series of events, *i.e.*, a given distribution of wave packets, one constructs a density matrix $\rho_e$ consisting of the tensor product of all the states $|G\rangle\langle G|$, which can have an interaction with *A* between times *t* and $t+\delta t$. A density matrix $\rho_{A+e}(t)$ is defined as $\rho_e(t) \otimes \rho_A(t)$. It evolves because of collisions as discussed earlier and $\rho_A(t+\delta t)$ is obtained according to Eq.(2.1) as the trace over $e$ of $\rho_{A+m}(t+\delta t)$.

Injection of incoherence into a system through its interaction with an environment is a universal phenomenon. A spectacular consequence is decoherence, which consists essentially of different behaviors in different channels when there is measurement. One may begin its study in the case when there is no measurement, or rather the case of a measuring apparatus before measurement.

The first term in the right-hand side of Eq.(3.12) describes a sector $|k\rangle$ losing a part of its norm from losses towards other sectors under the effect of individual collisions. There is no effect of a random phase in that term. The second one brings into Sector $k$ incoherent gains from other sectors. The gain and loss in probabilities from the two effects do not compensate in an individual sector $|k\rangle$, but they do so on average over all the sectors since the total gains and losses are equal, as shown in Eq.(3.5).

Because of the collisions during a time interval $\delta t$, the perturbed wave function of a sector $k$ becomes

$$e^{i\alpha_k}|k\rangle \to a_k e^{i\alpha_k}|k\rangle + b_k|k\rangle, \qquad (3.16)$$

where $a$ denotes a positive number smaller than 1 and $b$ a sum of small incoherent random numbers. Using Eq.(3.11), one gets on average $\delta a_k \leq 0$ and

$$\langle \delta a_k^2 \rangle = \langle \delta |b_k|^2 \rangle = \Phi\sigma\delta t, \qquad (3.17)$$

where $\Phi$ is the flux of incoming particles and $\sigma$ their total collision cross section (often equal in practice to the area of a boundary of $A$).

In the case of a two-channel measurement, one may restrict the discussion to an example where $A$ is only a pointer with position $X$ [10]. Because of the large mass of the pointer, a collision with an external particle cannot change its position and $X$ behaves like a conserved quantity. Two state vectors $|A,k1\rangle$ and $|A,k2\rangle$ (analogous to $|A1\rangle$ and $|A2\rangle$ in Eq.(1.2)) interact independently with a particle $M$: because the corresponding $T$-matrix elements are out of phase when the two positions of the pointer are sufficiently distant. This independence is realized when

$$|x_1 - x_2| \gg \hbar/|q|. \qquad (3.18)$$

(See [10] for a derivation of this property from the space-translation invariance of the $S$-matrix and for a consideration of the case when $x_1 - x_2$ is smaller than the de Broglie wavelength $\hbar/|q|$ of $M$).

One is interested in the reduced density matrix involving the values of $X$ and one may write for illustration one diagonal element and one non-diagonal element of $\rho_r$, namely

$$\rho_r(x_1, x_1) = |c_1|^2 \sum_k p_k \langle Ak1|Ak1\rangle \qquad (3.19a)$$

$$\rho_r(x_1, x_2) = c_1 c_2^* \sum_k p_k \langle Ak2|Ak1\rangle. \qquad (3.19b)$$

Before measurement, the two states $|Ak1\rangle$ and $|Ak2\rangle$ coincide in the density matrix $\rho_A(t)$ and their phases $\alpha_k$ are identical. On average over all sectors, the diagonal term.(3.19a) is invariant, because there are no quantum transitions between the two channels in view of the conservation of *X*. The sum (3.19a) in the diagonal term is therefore conserved through collisions.

In the case of the non-diagonal term (3.19b), the matrix elements $\langle Ak1,q|T|Ak'1,q'\rangle$ and $\langle Ak2,q|T|Ak'2,q'\rangle$ are out of phase if $x_1 - x_2$ is large enough. The incoherent quantities $b_k$ in Eq.(3.16) give a vanishing contribution to the scalar products in Eq.(3.19b) and one gets

$$\langle Ak1,t|Ak2t\rangle = a_{k1}.a_{k2}; \qquad (3.20)$$

or, in view of Eq.(3.17),

$$\rho_r(x_1,x_2;t) = \rho_r(x_1,x_2;0)\exp[-\Phi\sigma t] \qquad (3.21)$$

These conclusions coincide with the standard ones. They should have done so in any case, because one can see that the present ones do not depend strongly on the interpretation of *e*. The notion of a probabilistic environment was used because of its convenience but the traces over the state of *e* would have given exactly the same results in a model where *e* is considered as a quantum system. A different question is however whether some other effects can occur in *A* "itself", in which case they would also require an extraction of $\rho_{A+m}$ from $\rho_{A+m+e}$, whatever the model for *e*. This is the topic of the next section.

## 4. Fluctuations in probabilities

In spite of the wealth of its results and its perspectives [13, 30], decoherence theory remains often limited by an insistence on the role of pointers and of the reduced density matrix. Its interpretation as a growth of entanglement ignores moreover that an injection of incoherence from an external environment holds permanently, almost everywhere, irrespectively of measurements. Two notions are useful for extending these perspectives and they are concerned with the locality of incoherence injection into a macroscopic system and with the organization of a real experimental device. One will first look at them:

*Local generation and propagation of incoherence*

The collisions generating decoherence are local. An external molecule hits the apparatus in a limited region. The resulting local perturbation of wave functions expands from this initial region to wider ones in *A*. Its propagation is not instantaneous and does not bring out a sudden global transition $|kq\rangle \to |k'q'\rangle$ between different sectors of $\rho_A$, as was described in Section 3. This propagation is well known from the standpoint of kinetic theory but is much involved in its quantum aspects. There are two ways for bridging this gap. One of them is a localization of quantum collision theory using systematically wave packets, but it is non-trivial (see[28], chapters 8, 9, 11, and [29]). These works recover several classical notions regarding locality such as time delays and a finite propagation velocity of a local perturbation of wave functions, which proceeds through collisions at the atomic scale, often at the velocity of sound.

Another link between quantum and classical concepts relies on the use of a Wigner function to express a density matrix [31], and the associated idea of a Weyl symbol for an operator [32]. They both found their full expression in microlocal analysis [33, 34], which allowed a convenient derivation of classical dynamics [12]. One can sketch this approach as follows: One considers a system made of atoms with phase-space coordinates $(x_k, p_k)$. The Weyl-symbol of the Hamiltonian is the classical Hamilton function depending on these variables and is denoted presently by $H$. The Wigner function $W$ is the Weyl-symbol of $\rho$. Introducing the differential operator

$$\sigma \equiv \sum_k \left( \frac{\overleftarrow{\partial}}{\partial x_k} \cdot \frac{\overrightarrow{\partial}}{\partial p_k} - \frac{\overleftarrow{\partial}}{\partial p_k} \cdot \frac{\overrightarrow{\partial}}{\partial x_k} \right), \tag{4.1}$$

and $\quad \circ \equiv \exp(-i\hbar\sigma/2),$ (4.2)

the evolution of $W$ is given by

$$\partial W / \partial t = (1/i\hbar)(H \circ W - W \circ H) + \partial(\delta W)/\partial t, \tag{4.3}$$

where $\delta W$ represents the injection of incoherence. This effect is random and localized initially near the boundary of $A$. As shown in Section 3, one has $Tr\,\delta\rho = 0$ or equivalently

$$\int \delta W\, dx\, dp \equiv \int \delta W \prod_k dx_k\, dp_k (2\pi\hbar)^{-3} = 0. \tag{4.4}$$

As also shown in Section 3, $\delta\rho$ consists of gains and losses (a positive and a negative part), which can be associated though not exactly (in view of Gårding's sharp inequality ([33] and [12], Chapter 6) to a positive and a negative contribution to $\delta W$. In spite of their overall compensation in Eq.(4.4), these two terms are significantly significant. The positive part, for instance, brings the momentum resulting from external pressure into $A$ and the negative part carries no momentum. The compensation for this injection of momentum accompanying injection of incoherence results from the first term in the right-hand side of Eq.(4.3), which describes standard quantum mechanics and implies diffusion.

The leading term in $\hbar$ of this term yields the classical equation involving a Poisson bracket for the evolution of $W$ together with a strictly quantum correction, namely

$$\partial W / \partial t = \{H, W\} + O(\hbar^2). \tag{4.5}$$

This equation provides a convenient approximate expression for the propagation of incoherence, because the leading term represents a classical propagation with velocities $p_k/m$ whereas the terms of higher order in $\hbar$ are local (at least for a finite range of the potential interaction between atoms). These higher-order terms do not affect therefore the existence of propagation though they can affect significantly its detailed features.

When a further approximation is performed, a diffusion term $-D\sum_k (\partial/\partial x_k)^2$ can be used in place of the Poisson bracket in Eq.(4.5) and the resulting diffusion equation can yield a convenient estimate for some kinetic properties. There are also similar diffusion effects in momentum,, often described by a Fokker-Planck equation [35]. Together with spatial diffusion, they bring in the momentum transfer compensating momentum injection from outside. The present description is of course very sketchy, but it indicates a high complexity in the phenomena occurring in the transport of incoherence.

Finally, as a word of caution, one may recall that the localization of an atom is rather straightforward in the Wigner function, but it is much more involved in wave functions. This is because the density matrix is a Fourier transform of $W$ and the wave functions, which are its eigenfunctions, depend on the values of the density matrix everywhere. Fortunately, the properties with which one deals in this section will be only concerned with the localization of some atoms inside a rather large region of space, and there is no problem in this case..

*Organization of an apparatus*

Another useful item in a study of collapse is the organization of a real apparatus. This concept of organization is not meant presently as self-organization, which holds for instance in a solid part of an apparatus [36]. Its present meaning is more commonplace and refers to the fact that a measuring apparatus consists of different parts, which were conceived, manufactured and joined together to make them act collectively as a measuring device. A quantum formalism describing this kind of organization has been devised for this purpose [37], but it will not be needed here and one will rather use a simple model of $A$ where only two parts are distinguished. There is a reactive part $R$ in which macroscopic signals grow after an initial interaction between $A$ and $m$ (these signals can consist for instance of ionized tracks in a Geiger counter or in a wire chamber, tracks of excited atoms in a Cerenkov detector, droplets, bubbles, dislocations,…). The reactive component can also include a pointer, after it reacted to the measurement and moved over a finite distance. There are many examples, but their common feature is that $R$ carries different macroscopic indicators (different tracks, either track or no track, different positions of a pointer) in distinct measurement channels, and these indicators are conserved under collisions with the environment (as was the case for $X$ in the previous section) .

Besides the reactive part of $A$, one introduces a background $B$, which is its passive part. $B$ is often substantial in an actual apparatus where it consists in some sense of "everything else", except the core of measurement. It includes for instance some parts of the dielectric outside the tracks in a Geiger counter, or the battery producing an electric field. $B$ can even include a pointer before it began to move while minute signals occurred already in $R$. The apparatus $A$ is thus considered as a combined system $R + B$.

When organization and locality are put together, one expects that a measurement begins by a break in the state of the reactive region into states $R_1$ and $R_2$ carrying different signals. Then this effect propagates and acts also on the background, splitting its unique initial state into distinct states $B_1$ and $B_2$. Finally, in an ideally isolated apparatus, the sectors of $\rho_{A+m}$ show an entanglement

$$c_1|B_1\rangle|R_1\rangle|1\rangle + c_2|B_2\rangle|R_2\rangle|2\rangle \qquad (4..6)$$

This refined description of a measurement boils down essentially to the usual one and it brings nothing new until interaction with the environment is introduced. No change in the decoherenece effect results from this introduction and the non-diagonal parts in $\rho_{A+m}$ still tend to vanish. On the other hand, the previous conservation of the channel probabilities $p_1$ and $p_2$ must be reconsidered. In Section 3, like in usual decoherence theory, this conservation was based on the conservation of signals (or of the pointer positions) and the apparatus was either considered as essentially indivisible or conceptually reduced to a pointer. When one looks at Eq.(4.6) however, one sees again that the two states $R_1$ and $R_2$ carry different signals so that no external collision can create a quantum transition between them. But this is not true for the states $B_1$ and $B_2$ between which there is no "superselection" rule. A quantum transition between $B_1$ and $B_2$, producing a change in their respective norms, would yield also a change in the respective probabilities of the two pointer positions.

A possibility of fluctuations in channel probabilities suggests immediately a wider one, which would be a complete collapse resulting from an accumulation of fluctuations [38]. It would be quite new since, if so, collapse would occur in the passive part of the apparatus and not in the active part (including the pointer) on which attention has been always concentrated. This remarkable possibility will be now investigated in a specific model, although with a slightly different and somewhat subtler result: One will find that a process of collapse would not be exactly the one that was just mentioned and fluctuations in probabilities do not from a direct exchange between $B_1$ and $B_2$: They can do so earlier, when the state of $B$ is not yet split completely into $B_1$ and $B_2$ but is still in the course of splitting.

*A model*

The model to be considered is essentially Schrödinger's initial example, because of its simplicity: It consists of a radioactive nucleus in a Geiger counter. One studies the behavior of this counter during some time interval $\Delta T$ so that the two states $|1\rangle$ and $|2\rangle$, introduced in Section 1, stand for the states of the nucleus after decay or with no decay. To avoid an elaborate description of organization, one will consider the origin of the electric field in the counter as external and with no connection with the environment producing decoherence

When the nucleus decays, it emits an $\alpha$-particle with an isotropic wave function. The observed existence of linear tracks is however probably as important for interpretation as this isotropic emission. Mott studied the consistency of these two aspects long ago and showed essentially that the $\alpha$-emission consists in an isotropic quantum distribution of radial tracks [39] This question was also reconsidered in the framework of consistent histories [12, Chapter 5] and, when applied to the present case, this approach shows that the correlation of ions along a track is unrelated to the state of the medium and depends only on the kinetic Hamiltonian of the $\alpha$ particle. This means that the existence of tracks, which is a quantum effect, is insensitive to the flux of incoherence resulting from external collisions.

To set the stage, it will be convenient to start as in Section 1 from the reliability conditions. If the state $|1\rangle$ denotes the case when a decay occurred, this event is certain if the lifetime of the radioactive nucleus is very short. Along a track consisting of primary electrons and ions, secondary ionizations are prduced rapidly because of acceleration of the primary electrons by the electric field. There is cascade of increasing ionization, which reaches a macroscopic level until a spark occurs.

If there is no electric field however, everything remains at atomic level. Primary electrons disappear some time later through recombination and nothing remains of the decay. The electric field yields therefore the necessary communication between atomic and macroscopic levels, between quantum indeterminism and a unique reality. This communication, or transition, goes through the mediation of growing signals (a growing ionization in the present case). The irreversible character of this growth is certainly an essential character, but one will not search in that direction because irreversible thermodynamics is not well enough understood in its foundations. One will only assume as necessary for a measurement the occurrence of macroscopic effects originating in the measured microscopic event, *i.e.*, in decay in the present case.

Assuming the range of the $\alpha$-particle smaller than the size of the detector, one defines then the active part $R$ of $A$ as the sphere containing possible initial tracks while the background $B$ consists of the dielectric (supposed to be a gas) outside of $R$ and of the boundary on which external molecules collide. Still in the case of certainty for decay, the initial probability for an individual atom in $R$ to interact with a primary electron or ion is small, but it grows exponentially and reaches soon the value 1 when secondary ions are produced. There is nothing of that kind in *B* where every atom ignores essentially that a measurement has occurred until it is reached by propagation of the state in *R*.

When the lifetime of the radioactive nucleus is not very short, one must deal with a superposed state (1.2) for this nucleus, where $|2\rangle$ denotes the state of no-decay where nothing happens. The definition of the regions *R* and *B* remains the same. It refers only to the case of decay, but this is only an unimportant peculiarity of yes-no experiments (decay versus no decay). Because decoherence affects strongly the state of the whole apparatus *A*, one can make only guesses about the states of *R* and *B* before making a thorough analysis, which will be certainly difficult. Anyway, one assumes that, after a sufficient growth of secondary ionization, an instantaneous state of the atoms in *R* has become completely entangled with the two states of *m* (the nucleus) whereas the state in *B* ignores this entanglement. Rather than Eq.(4.6), this situation would be expressed by a factorized state

$$(c_1|R_1\rangle|1\rangle + c_2|R_2\rangle|2\rangle) \otimes |B\rangle. \quad (4.7)$$

It will be interesting to describe also this expression (which is only approximate) in terms of density matrices. To begin with, one notices that the argument in Section 3 showing a very good diagonality for the reduced density matrix works as well for the full density matrix $\rho_{A+m}$. One has then with a good approximation

$$\rho_{A+m} = p_1 \rho_{A11} \otimes |1\rangle\langle 1| + p_2 \rho_{A22} \otimes |2\rangle\langle 2| \quad (4.8)$$

and in view of Eq.(4.7), one has with a rougher approximation

$$\rho_{Ajj} \approx \rho_{Rj} \otimes \rho_B, \quad (4.9)$$

where $\rho_B$ does not depend on the channel ($j = 1$ or $2$) under consideration.

It might be noticed incidentally that this use of approximate expressions goes against a frequent insistence on the idea that the uniqueness of reality is such a fundamental question that it cannot be approached through approximate methods, valid "for all practical purposes"[40]. This is because an emergence of uniqueness from quantum dynamics would be necessarily approximate, like the emergence of classical dynamics and as shown by the atomic properties of a real macroscopic object. One considers therefore this restriction as irrelevant.

The real meaning of Eq.(4.9) can be understood better with the help of Wigner functions: ($W_{Ajj} \approx W_{Rj} W_B$) where a detailed consideration of the phase space coordinates can refine the meaning of the rough expression (4.9). But this is only an incidental remark to signal a rather wide flexibility in the present approach and the corresponding difficulty in getting the best mathematical formulation for it.

*Probability fluctuations*

One arrives now at the crux of the problem, which is the possibility of fluctuations in $p_1$ and $p_2$ when ionization becomes macroscopic. Several points are worth noticing: There are many different potential tracks and each one of them could be considered as a different channel. In that case, collapse would have also to elect one of them. The state inside the apparently homogeneous sphere *R* is also therefore highly structured. It fluctuates, because of incoherent perturbations arriving in it; they were generated by the environment in different locations on the boundary and often a long time before measurement. This difference in the time and location of their origin is probably of little significance for the overall spherical region $R$, but it has certainly important effects among the channels showing different tracks.

Then there is the boundary between $R$ and $B$. It is obviously imprecise and its width and motion are certainly controlled by diffusion in a first approximation. Its shape depends also strongly on the channel-track under consideration and evolves. Finally, the standard description of diffusion by a diffusion equation is probably too rough and the forefront, namely the region where the diffusion equation predicts a non-vanishing though very small probability for an atom from $R$ to penetrate in $B$ is presumably significant. One can only predict with some confidence that, well inside $R$, there is complete entanglement. Similarly, well inside $B$, there is no entanglement.

It seems difficult to adapt no-go theorems to this unexplored situation and conservation of the channel probabilities $p_1$ and $p_2$ becomes problematic. A study of their possible variation is rather embarrassed on the other hand by too many opportunities for fluctuations and a difficulty in evaluating their global consequence. Nevertheless, one can point out an interesting mechanism, which though it has not yet been analyzed thoroughly in a quantitative way, seems promising or at least illustrative: One considers a region $\beta$ in $B$, somewhat far from the $R$-$B$ boundary but inside the forefront, namely such that the probability for an atom from $R$ to reach $\beta$ is small but non-vanishing at some specific time. This is true for both states $|1\rangle$ and $|2\rangle$ Only a few atoms arrive into $\beta$ from $R$, where entanglement is complete. Because they are so few, the fluctuations in their number are large whereas their entanglement with $m$ does not coincide with $p_1$ and $p_2$, the difference being a priori strongest when the number of exceptional incoming atoms is smallest.

One denotes by $q_1(0)$ and $q_2(0)$ the small probabilities for these few incoming atoms to carry respectively an entanglement with either $|1\rangle$ or $|2\rangle$. One denotes by 0 the time of their entry in $\beta$, and by $q_0(0)$ the initial probability, close to 1, for no entanglement in $\beta$. One has then $q_1(0) + q_2(0) = \varepsilon$ and $q_0(0) = 1 - \varepsilon$, with $\varepsilon$ a small number.. Every collision of a non-entangled atom with an atom entangled with $|1\rangle$, for instance, brings the two outgoing atoms into entanglement with $|1\rangle$ and this effect leads to a cascade growth of entanglement, expressed by the equations

$$dq_0/dt = -q_0(q_1 + q_2)/\tau, \; dq_1/dt = q_0 q_1/\tau, \; dq_2/dt = q_0 q_2/\tau, \quad (4.11)$$

where the parameter $\tau$ is the mean free collision time of atoms. This gives after integration

$$q_0(t) = [(1-\varepsilon)\exp(-t/\tau)]/[\varepsilon + (1-\varepsilon)\exp(-t/\tau)],$$
$$q_1(t) = \delta q_1/[\varepsilon + (1-\varepsilon)\exp(-t/\tau)], \quad (4.12)$$
$$q_2(t) = \delta q_2/[\varepsilon + (1-\varepsilon)\exp(-t/\tau)] \; ,$$

with asymptotic values

$$q_1 \to \delta q_1/\varepsilon, q_2 \to \delta q_2/\varepsilon., \; q_0 \to 0. \quad (4.13)$$

Initial fluctuations are therefore strongly amplified through this process.

Near the forefront of the frontier between $R$ and $B$, there are many small regions $\beta$ (the frontier itself can be shown to move with a velocity $v = 2(D/\tau)^{1/2}$ during the growth of $R$ ). Another approximation, borrowed from statistical mechanics , consists then in introducing the $\beta$ regions as so many Gibbs sub-ensembles having the same size and writing in place of Eq.(4.8-9):

$$\rho_{A+m} = p_1 \rho_{R1} \otimes \prod_\beta \rho_{\beta 1} \otimes |1\rangle\langle 1| + p_2 \rho_{R2} \otimes \prod_\beta \rho_{\beta 2} \otimes |2\rangle\langle 2| \qquad (4.14)$$

(this kind of factorization, which is approximate, is known to yield an additive entropy [41]). The previous factorization in Eq.(4.9) means then that $\rho_{\beta 1} = \rho_{\beta 2}$ for a Gibbs subsystem $\beta$ in $B$ that is far enough from the frontier with $R$. One denotes the first two asymptotic values(4.13) by $\delta p_{\beta 1}$ and $\delta p_{\beta 2}$ for some $\beta$ and one introduces the time interval $\delta t$ during which the frontier between $R$ and $B$ crosses a $\beta$ region. One gets then total fluctuations for $p_1$ and $p_2$ during this time $\delta t$, which is given by

$$\delta p_j = \sum_\beta \delta p_{\beta j}, \qquad (j = 1 \text{ or } 2) \qquad (4.15)$$

the summation being made on the Gibbs subsystems in the forefront of the $R$-$B$ boundary.

The various local fluctuations $\delta p_{\beta j}$ are independent, except for the constraint $\delta p_1 + \delta p_2 = 0$ on their sum, which results from the exact relation $p_1 + p_2 = 1$ expressing the unit trace of $\rho_{A+m}$. The correlation coefficient $A_{12} = \langle \delta p_1 \delta p_2 \rangle$ is therefore negative and one gets with the same notation $A_{11} = A_{22} = -A_{12}$. A computation of the expression and of the value of these coefficients is non-trivial however, as well as an assessment of these rough considerations on a firmer ground. The best one can do is to guess that $A_{12}$ is presumably proportional to the product $p_1 p_2$ if, for instance, $q_{\beta 1}(0)$ is roughly proportional to $p_1^{1/2}$ as one would expect of the fluctuation of a random small sample of atoms getting out of a box $R$ in which their proportion is $p_1$. An especially difficult problem is to take care of the structure of $R$ in linear tracks, not even mentioning their competition and the resulting fluctuations in the probabilities of different tracks.

It could be that these simple remarks contain in germ a theory of collapse but the present exploration contains much guesswork. A more advisable conclusion at the present stage of investigation would be therefore to say:: "Fluctuations in $p_1$ and $p_2$ are not impossible and one cannot reject a priori that they could lead to an emergence of collapse from quantum dynamics".

**5. From fluctuations to reduction**

In the case of two channels ($j, j'$) with $j, j'$ equal to 1 or 2, one has defined correlations coefficients $A_{jj'} = \langle \delta p_j \delta p_{j'} \rangle$ for the fluctuations occurring during a short time interval $\delta t$. One has then

$$A_{11} = A_{22} = -A_{12}, \qquad (5.1)$$

because $\delta p_1 + \delta p_2$ is exactly zero, and this implies $A_{12} \leq 0$.

The question whether fluctuations in channel probabilities can yield reduction was solved by Pearle long ago [38]. He assumed an origin of fluctuations in nonlinear violations of the Schrödinger equation, but it was shown here that they could also be due to an environment with no violation of the basic axioms of quantum mechanics. Pearle's essential conclusions can be summarized by a theorem dealing with an arbitrary number of channels, denoted by an index *j*. This theorem relies on the following assumptions:

*Assumption 1:* The probabilities $p_j$ of the various channels evolve randomly (one may conveniently call this variation of the $p_j$'s a Brownian motion).

*Assumption 2:* The correlation functions $A_{jj'}$ of the fluctuations occurring during a short time $\delta t$ are proportional to $\delta t$, as in a Brownian process. They depend only on time and on the $p_j$'s themselves.

*Assumption 3:* If some probability happens to vanish during the Brownian motion, it remains zero afterwards.

According to Pearle's theorem, some channel probability $p_j$ must inevitably become equal to 1 after some time and the occurrence of this final event is random. The other probabilities vanish then and, at the end, the density matrix has become

$$\rho_{A+m} = \rho_{Aj} \otimes |j\rangle\langle j|, \qquad (5.2)$$

and the position of the pointer in $\rho_{Aj}$ is $x_j$. This result agrees perfectly with wave function collapse when the measured system $m$ is not destroyed and its state is not modified during measurement. The extension to more general situations is straightforward.

The seminal result of Pearle's theorem lies moreover in the predicted probabilities for the possible outcomes of the Brownian process: This Brownian probability for the *j*-outcome is exactly equal to the initial value of $p_j$ before the beginning of fluctuations, namely: $p_j = |c_j|^2$. Pearle's theorem implies therefore that Born's rule, usually considered as a principle of quantum mechanics, could be actually a consequence of the other principles.

For use in the next section, one may add that the proof of this theorem relies on a Fokker-Plank equation for a Brownian probability distribution $Q(p_1, p_2, p_3, ...)$ expressing the random variations of the quantities $p_k$, which are considered as variables. One has

$$\partial Q / \partial t = \sum_{jj'} \partial^2 (A_{jj'} Q) / \partial p_j \partial p_{j'}. \qquad (5.3)$$

One can also show that reduction occurs after a random time having an exponential distribution $\exp(-t/\tau_{red})$ for large values of $t$.

Pearle's assumptions 1 and 2 are presumably valid in the present theory. Assumption 3 is certainly valid, since it means that when a channel disappeared together with its own macroscopic signal from a reactive part $R$ of the apparatus, no internal effect occurring in the passive part $B$ (the locus of fluctuations) could recreate in $R$ this lost signal

**6. Non-separability**

Quantum mechanics is non-separable [1], and this remarkable property was confirmed by measurements of an Einstein-Podolsky-Rosen pair of photons by two space-like separated apparatuses [42-44]. Collapse was described here however as a local effect occurring in a unique apparatus $A$ and the existence of a correlation between two distant apparatuses raises therefore a worrying problem, which was pointed out privately by Nick Herbert and by Bernard d'Espagnat (who was also very helpful in a detailed check of its answer in the present section). One cannot ignore this question in a paper dealing with matters of principle and one must therefore investigate how the present theory using local reduction can nevertheless agree with these experimental results.

One considers for convenience the case of two spin-1/2 particles 1 and 2, because the mathematics is the same as in the case of polarized photons and slightly easier. The eigenvectors of the *z*-components of two spins are supposed entangled in a normalized initial state

$$|\psi\rangle = a|1z,+\rangle|2z,-\rangle + b|1z,-\rangle|2z,+\rangle . \quad (6.1)$$

Here, $|1z,+\rangle$ for instance, denotes the normalized state of particle 1 with its *z*-component of spin equal to +1/2, other notations being similar. Two space-like separated apparatuses $A_1$ and $A_2$ measure respectively the components of spin 1 and spin 2 along another space direction *z'* making an angle $\theta$ with the direction *z*. When expressed in the *z'*-basis, the state (6.1) becomes

$$|\psi\rangle = -(a+b)cs|1z',+\rangle|2z',+\rangle + (ac^2 - bs^2)|1z',+\rangle|2z',-\rangle$$
$$+ (bc^2 - as^2)|1z',-\rangle|2z',+\rangle + (a+b)cs|1z',-\rangle|2z',-\rangle, \quad (6.2)$$

with $c = \cos(\vartheta/2)$, $s = \cos(\vartheta/2)$. It will be convenient to write down the four vectors occurring in this expression as $|\alpha,\beta\rangle$ where the labels $\alpha$ and $\beta$ take the two values ± and, for instance, $|++\rangle$ stands for $|1z',+\rangle|2z',+\rangle$. One has then

$$|\psi\rangle = \sum_{\alpha\beta} c_{\alpha\beta}|\alpha\beta\rangle. \quad (6.3)$$

Experiments show that the observation of the result $\alpha$ by $A_1$ and the result $\beta$ by $A_2$ has the frequency $p_{\alpha\beta} = |c_{\alpha\beta}|^2$. This means that the correlation coefficients are insensitive to the localization of measurements and is a clean manifestation of quantum non-separabilty .

One chooses a space-time reference frame in which the two measurements begin at the same time. Just after detection, small signals occur in the reactive regions $R_1$ and $R_2$ of the apparatuses $A_1$ and $A_2$. The quantum probability for the signal $\alpha$ in $R_1$ and simultaneously the signal $\beta$ in $R_2$ is $p_{\alpha\beta}$, according to the definition of quantum probabilities as squares of amplitudes. For more clarity, it will be convenient to introduce a language (which could have been also previously useful) in which squares of quantum amplitudes are called "quantum weights", making clearer that they can evolve randomly under dynamical effects in the passive regions $B_1$ and $B_2$. One then says that $p_{\alpha\beta}$ is the quantum weight for the signals $\alpha$ in $R_1$ and $\beta$ in $R_2$, before any splitting in the states of the passive regions $B_1$ and $B_2$.

Some time later, the states of both $B_1$ and $B_2$ split also and begin to show fluctuations. The "Brownian" probability for a signal $\alpha$ in $A_1$ and $\beta$ in $A_2$ is no more equal to $p_{\alpha\beta}$. It fluctuates and one denotes it by $q_{\alpha\beta}(t)$. In the Fokker-Planck equation (5.3), which expresses only the existence of fluctuations without any reference to their origin, the variables in the Brownian probability distribution $Q$ are now the numbers $q_{\alpha\beta}$, which one writes $q_j$ after introducing again a change in notation $\alpha\beta \to j$. Eq.(4.3) becomes then

$$\partial Q/\partial t = \sum_{jj'} \partial^2 (A_{jj'} Q)/\partial q_j \partial q_{j'}, \qquad (6.4)$$

with

$$A_{jj'}(t) \equiv A_{\alpha\beta,\alpha'\beta'}(t) = \langle \delta q_{\alpha\beta}(t) \delta q_{\alpha'\beta'}(t) \rangle. \qquad (6.5)$$

One may come now to the central question arising from the locality of fluctuations in $B_1$ and $B_2$. This means that a fluctuation in $B_1$ does not produce a fluctuation in $B_2$, and conversely. This independence implies that, during a short time $\delta t$, one has

$$\delta q_{\alpha\beta} = \delta q_\alpha^{(1)} + \delta q_\beta^{(2)}, \qquad (6.6)$$

where the two fluctuations in the right-hand side occur respectively in $B_1$ and $B_2$. They are independent, so that

$$\langle \delta q_\alpha^{(1)} \delta q_\beta^{(2)} \rangle = 0. \qquad (6.7)$$

One has therefore

$$A_{\alpha\beta,\alpha'\beta'} = A_{\alpha\alpha'}^{(1)} \delta_{\beta\beta'} + A_{\beta\beta'}^{(2)} \delta_{\alpha\alpha'}, \qquad (6.8)$$

where $A_{\alpha\alpha'}^{(1)}$ for instance is a Brownian correlation coefficient for the fluctuations in $B_1$. This special form of correlation coefficients for the joint apparatuses does not spoil the assumptions of Pearle's theorem and the conclusions remain valid, namely a unique pure state of the spins $|\alpha\beta'\rangle\langle\alpha\beta'|$ will finally come out, associated with the corresponding positions of pointers in $A_1$ and $A_2$. The corresponding Brownian probability will be the initial value of $q_{\alpha\beta}$, namely $p_{\alpha\beta}$, and this is in full agreement with experiments. A similar analysis can be made when the two measurements are performed successively and also when the second one begins before a complete reduction of the first one, with identical results.

## 6. Conclusions

Several points regarding decoherence and collapse were brought to attention in this work:

- An environment cannot be considered as a properly defined quantum system if one is not disposed to bring up the whole universe to explain a standard effect at non-cosmological scale.
- A probabilistic description of environment points out a lack of unitarity in the internal evolution of an open system.
- Decoherence appears in this framework as a propagating effect, which occurs as an injection of incoherence, even in a non-measuring macroscopic system.

- Organization can have a significant role in the performance of a measuring apparatus, particularly through a distinction between active and passive parts, probably more significant than the usual concentrating upon the notion of pointer.
- The passive part, in which no conserved quantity takes different values, could be the locus of probability fluctuations leading ultimately to collapse.
- Such a theory of collapse would not conflict with the non-separable character of quantum mechanics.

These remarks raise altogether a strong suspicion against the allegations of inconsistency between collapse and the quantum principles, previously considered as no-go theorems. They unlock a wide range of phenomena, which could not investigated in detail in the present work because of their complexity. Some points regarding matters of principle were sketched however. They look promising enough for supporting more thorough research.

## Acknowledgements

The construction of this theory through many misgivings and mistakes would not have been possible without liberal and pertinent criticisms and remarks from several colleagues, particularly Bernard d'Espagnat and Heinz-Dieter Zeh. I thank also Robert Dautray, Bernard d'Espagnat, Jacques Friedel, Frank Laloë and Philip Pearle for their encouragement.